\documentclass[12pt,a4paper]{article}

\usepackage{epsfig}
\usepackage{exscale}
\usepackage{color}
\def\lsim{\raise0.3ex\hbox{$<$\kern-0.75em\raise-1.1ex\hbox{$\sim$}}}
\def\gsim{\raise0.3ex\hbox{$>$\kern-0.75em\raise-1.1ex\hbox{$\sim$}}}
\newcommand{\be}{\begin{equation}}
\newcommand{\ee}{\end{equation}}
\newcommand{\ba}{\begin{eqnarray}}
\newcommand{\ea}{\end{eqnarray}}

\def\spose#1{\hbox to 0pt{#1\hss}}
\def\ltapprox{\mathrel{\spose{\lower 3pt\hbox{$\mathchar"218$}}
 \raise 2.0pt\hbox{$\mathchar"13C$}}}
\def\gtapprox{\mathrel{\spose{\lower 3pt\hbox{$\mathchar"218$}}
 \raise 2.0pt\hbox{$\mathchar"13E$}}}
\def\phv{\vec \phi}

\def\NT{N_\tau}

\def\nt{\ifmmode\NT\else$\NT$\fi}
\def\NS{N_\sigma}
\def\ns{\ifmmode\NS\else$\NS$\fi}

\def\p{^\prime}
\def\v{\vec}
\def\n{\noindent}

\begin{document}
\begin{titlepage} 
\thispagestyle{empty}

\mbox{} \hfill BI-TP 2014/04\\
\mbox{} \hfill arXiv:1402.5302v2\\
\mbox{} \hfill June 2014
\begin{center}
\vspace*{0.8cm}
{{\Large \bf Finite size dependence of scaling functions of the\\
three-dimensional \boldmath$O(4)$ model in an external field\\}}
\vspace*{1.0cm}
{\large J.\ Engels$^{\rm a}$ and F.\ Karsch$^{\rm a,b}$}\\ \vspace*{0.8cm}
\centerline{$^{\rm a}$ Fakult\"at f\"ur Physik, 
Universit\"at Bielefeld, D-33615 Bielefeld, Germany}
 \vspace*{0.3cm}
\centerline{$^{\rm b}$ Physics Department Brookhaven 
National Laboratory, Upton, NY 11973}
\vspace*{0.9cm}
{\bf   Abstract   \\ } \end{center}
\indent
We calculate universal finite-size scaling functions for the order 
parameter and the longitudinal susceptibility of the three-dimensional
$O(4)$ model. The phase transition of this model is supposed to be in 
the same universality class as the chiral transition of two-flavor QCD.
The scaling functions serve as a testing device for QCD simulations on 
small lattices, where, for example, pseudocritical temperatures are 
difficult to determine. In addition, we have improved the 
infinite-volume limit parametrization of the scaling functions by using
newly generated high statistics data for the $3d$ $O(4)$ model in the high
temperature region on an $L=120$ lattice.

\vfill \begin{flushleft} 
\noindent{\rule[-.3cm]{5cm}{.02cm}} \\
\vspace*{0.2cm}
E-mail addresses: engels@physik.uni-bielefeld.de, 
karsch@physik.uni-bielefeld.de, karsch@bnl.gov 
\end{flushleft} 
\end{titlepage}


\section{Introduction}
\label{section:intro}

The three-dimensional $O(4)$ spin model plays an important role for
our understanding of the low energy limit of Quantumchromodynamics (QCD) 
as well as its phase structure at non-zero temperature. The almost massless 
pseudo-scalar particles (pions) of QCD have been considered already early
as Goldstone particles, which receive a non-zero mass only due to the 
presence of a chiral symmetry breaking field, i.e. the non-zero light quark
masses. The consequences of this idea for the QCD phase 
diagram have been worked out in detail in a paper by Pisarski and Wilczek 
\cite{Pisarski}. Their conclusion is that, if QCD undergoes a second-order 
chiral transition in the limit of vanishing up and down quark masses, 
this transition is expected to belong to the universality class of the 
three-dimensional $O(4)$ spin model \footnote{This is not settled with 
certainty as the influence of the axial anomaly on the transition is still 
not fully understood. The effective restoration of $U_A(1)$ at high 
temperature may indeed trigger a first order transition.}. Thermodynamics 
in the vicinity of the QCD transition will then show universal features 
that are described by $O(4)$ scaling functions. In fact, recent studies 
of the quark mass dependence of the chiral transition in lattice QCD 
using the staggered fermion discretization scheme led to good agreement 
with infinite-volume $O(2)$ or $O(4)$ scaling functions \cite{MEoS}. 

Lattice QCD calculations with light dynamical quarks are performed in 
relatively small volumes. This is in particular the case for the 
computationally demanding chiral fermion formulations, i.e., for studies
of QCD thermodynamics performed with domain wall fermions \cite{hotQCDDWF}
or overlap fermions \cite{overlap}. A finite volume limits the
correlation length and thus, similar to the influence of an external field, 
it modifies the universal behavior in the vicinity of a second-order phase 
transition. It has been shown that the finite volume dependence of the 
chiral condensate close to the QCD transition temperature may be understood 
in terms of finite-volume scaling functions of the $3d$ $O(4)$ spin 
model \cite{Mendes}. 

 Calculations performed within the framework of an $O(4)$ symmetric 
quark-meson model, using an approximation scheme based on the functional 
renormalization group (FRG) approach \cite{Braun,Braun:2011iz}, suggest
as well that finite size effects may influence the determination of 
the chiral transition temperature, or more generally, the temperature
dependence of the chiral susceptibility in QCD. It thus is of interest
to arrive at quantitative results for the finite size dependence of the
scaling functions of the $3d$ $O(4)$ universality class which can be
used in scaling studies of thermodynamic observables determined
in lattice QCD. Within the FRG approach \cite{BraunO4} these scaling 
functions have been derived approximately (In Ref. \cite{BraunO4} the 
exponent $\eta$ is zero, unlike in the $O(4)$ class.). 

In a recent paper \cite{Engels2}, we had derived representations 
of the scaling functions of the $3d$ $O(4)$ model in the thermodynamic
limit. This was done in terms of expansions of the scaling functions
in the scaling variable $z=\bar t/h^{1/\Delta}$, where 
$\Delta=\beta\delta$ is the so-called gap exponent and $h=H/H_0$ and
$\bar t= (T-T_c)/T_0$ are the reduced field and temperature. The 
parameters of the expansions were deduced exclusively from Monte Carlo
data with finite external fields. In the paper \cite{Engels2} we 
discussed in detail the relations, the common features and the 
differences of the new parametrization to the existing, previous ones,
which nearly all used the Widom-Griffiths form 
\cite{Widom:1965,Griffiths:1967}, where both the scaling function and
the scaling variable depend on the magnetization $M\,$. We had also 
derived in Ref.\ \cite{Engels2} the corresponding parametrization of
the scaling function $f_f(z)$ for the singular part of the free energy
density. The scaling functions $f_G(z)$ and $f_{\chi}(z)$ of the 
magnetization and the susceptibility are obtained from $f_f(z)$ by 
taking appropriate derivatives with respect to $H$.

In this paper we will determine finite-volume scaling functions 
of the order parameter and the susceptibility directly from high 
statistics Monte Carlo simulations of the $3d$ $O(4)$ spin 
model for varying $L$ and finite external field. The paper is thus 
a direct extension of Ref.\ \cite{Engels2} to finite $L$ for a 
limited region of $z$. For further details which are not dependent 
on $L$ being finite one should therefore consult Ref.\ \cite{Engels2}.
As far as we know, the only other study of finite-size scaling
relations in the presence of an external field for the $3d$ $O(4)$
spin model is that of Ref.\ \cite{Mendes}. In that paper, however,
the finite-size scaling relations were examined only for two fixed 
values of $z$, namely for $z=0$ and $z=z_p$, that is at the critical 
temperature and on the pseudocritical line, whereas we cover a larger
region in $z$.   

Our paper is organized as follows. In section \ref{section:fssf}
we extend the relations for the infinite-volume scaling functions to
the case of finite $L=V^{1/d}$. The simulations of the $3d$ $O(4)$ model
on lattices with $L=24-120$ and the resulting finite volume dependencies
of the scaling functions $f_G$ and $f_\chi$ are then discussed in 
section \ref{section:sio4}. In addition we improve the parametrization
of the infinite-volume scaling functions as given in Ref. 
\cite{Engels2}. We close with a summary and the conclusions.

\section{Finite size dependence of scaling functions}
\label{section:fssf}

In order to introduce the finite size dependence of the 
scaling functions we reconsider parts of chapter 2 of Ref. \cite{Engels2}.  
The scaling functions are derived from the reduced (containing a factor
${\beta=1/T}$ ) free energy density,
\be
f = -\frac{1}{V} \ln Z(T,H,L) = f_s(T,H,L) +f_{ns}(T,H,L) \ .
\ee
Here, we have split the free energy density in a singular term
$f_s$, responsible according to renormalization group (RG) theory for
critical behavior, and a regular or non-singular term $f_{ns}$. 
The scaling laws near the critical point are derived from the RG scaling
equation for $f_s$. The derivatives of $f_{ns}$ contribute regular terms
to the scaling laws, which apart from the cases of the energy density
and the specific heat (for $\alpha<0$) are sub-leading near the critical
point. The dependence on $L$ or better on $1/L$ can be
treated as a correction to leading scaling behavior, that is $l=L_0/L$
takes the r\^{o}le of an additional relevant scaling field with
exponent $y_L=1$. The RG equation becomes then
\be
f_s(u_t,u_h,l,u_4,\dots)= b^{-d}f_s(b^{y_t}u_t,b^{y_h}u_h,b\,l,
b^{y_4}u_4,\dots) \ .
\label{RGse}
\ee
Here, $b$ is a free positive scale factor and the usual, remaining
relevant scaling fields are $u_t=c_t t$, $u_h=c_h H$, where
$t=(T-T_c)/T_c$. The $c_t,c_h$ and $L_0$ are model-dependent
(positive) metric scale factors. In addition, there are infinitely 
many irrelevant scaling fields $u_j$ with negative exponents $y_j$.
By choosing $b=u_h^{-1/y_h}$ for $H>0$ one obtains from 
Eq.\ (\ref{RGse}) the form of scaling functions which we want to 
discuss here
\be
f_s(u_t,u_h,l,u_{j>3})=u_h^{d/y_h}f_s(u_tu_h^{-y_t/y_h},1,
l\,u_h^{-1/y_h},u_ju_h^{-y_j/y_h})~.
\label{RGu}
\ee
The dependence of $f_s$ on the irrelevant scaling fields leads to
the non-analytic corrections to the scaling functions and is here of 
the form
\be
u_ju_h^{-y_j/y_h} = u_j(c_hH)^{-y_j/y_h},
\label{irre}
\ee  
and not $\sim L^{y_j}$ as in the conventional approach where one 
chooses $b\sim L$. Hence, we only have a single variable, 
$lu_h^{-1/y_h}$, that depends on $L$ and all other variables are
$L$-independent. Close to the critical point, when $H$ is small, the
contributions of the irrelevant scaling fields become negligible, 
because the $y_j$ are negative. The singular term $f_s$ is then a
universal function of the scaling fields $u_t, u_h$ and $l$ only and
\be
f_s\;=\; (c_hH)^{d/y_h}\Psi_2(c_tc_h^{-y_t/y_h}tH^{-y_t/y_h},
c_h^{-1/y_h}l\,H^{-1/y_h})~,
\label{fspsi}
\ee
where $\Psi_2$ is again a universal function but in contrast to
the thermodynamic limit now of two arguments. 
By comparison with the infinite-volume scaling laws one derives
\be
y_t\;=\;1/\nu\;,\; y_h\;=\;1/\nu_c\;=\;\Delta/\nu\;,\;{\rm or~}\;
\Delta\;=\;y_h/y_t~,
\ee
and the known hyperscaling relations between the critical exponents.

Instead of working with the metric scale factors $c_t$ and $c_h$ one
introduces usually new temperature and field variables $\bar t=tT_c/T_0$
and $h=H/H_0$ in the thermodynamic limit. We stick to this tradition
also in the finite volume case. Correspondingly, the scaling functions 
of the observables which are derivatives of the free energy density 
will depend (see Eq. (\ref{fspsi})) on the two variables
\be
z=\bar t/h^{1/\Delta}~,\quad {\rm and} \quad z_L=l/h^{\nu_c}~,
\label{scvar}
\ee
and the thermodynamic limit is recovered for $z_L=0$ at finite $h$.
For example, the order parameter, or magnetization becomes then
\be
M = -\frac{\partial f}{\partial H} =  h^{1/\delta} f_{G}(z,z_L) \ .
\label{orpar}
\ee
In order to specify the scaling variables $z$ and $z_L$ in a given
model calculation one needs to fix the non-universal scales
$T_0$, $H_0$ and $L_0$. As already mentioned, the first two are fixed
by demanding in the infinite volume limit
\be
M(t=0) = h^{1/\delta}\;\; {\rm and} \;\; M(h=0)= (-t)^\beta \; .
\label{normalization}
\ee
This implies
\be
f_G(0,0)\; =\; 1~,\quad {\rm and}\quad f_G(z,0) {\raisebox{-1ex}{$
\stackrel{\displaystyle\longrightarrow}{\scriptstyle z 
\rightarrow -\infty}$}}(-z)^{\beta}~.
\label{normfg}
\ee
The scale $L_0$ is fixed by a third normalization condition. 
We choose 
\be
z_L=1 \;\; {\rm for}\;\;\frac{M(t=0)_{|L}}{M(t=0)_{|L=\infty}} =
f_G(0,1)= 0.874 \; .
\label{zLnorm}
\ee
That is, on a given lattice of size $V=L^d$ we determine at $t=0$ the 
finite-volume scaling function $f_G(0,z_L)$ by varying the external 
field $h$. We then assign the value $z_L=1$ to that value of 
$h$ that yields $f_G(0,1)=0.874$. This fixes the scale $L_0$ to be 
$L_0= L h^{\nu_c}$.
Our normalization condition seems arbitrary. In fact, we could have
chosen other normalization conditions. However, in the absence of any
obvious natural choice the above normalization condition is convenient.
As we shall see this condition leads to $L_0\equiv 1$ for the
$3d$ $O(4)$ spin model studied here.

Due to Eq. (\ref{fspsi}) and the derivative in Eq. (\ref{orpar}) the
singular term $f_s$ depends on the scaling function $f_f(z,z_L)$
\be
f_s = H_0h^{1+1/\delta}f_f(z,z_L) \ ,
\label{ffunc}
\ee
which relates to $f_G$ by
\be
f_{G} (z,z_L) = 
-\left(1+\frac{1}{\delta}\right) f_f(z,z_L)
+\frac{z}{\Delta}\frac{\partial f_f(z,z_L)}{\partial z}
+\nu_c z_L\frac{\partial f_f(z,z_L)}{\partial z_L} \ .
\label{fgff}
\ee

Fluctuations of the order parameter in the $O(4)$ model, are
described by the longitudinal susceptibility,
\be
\chi_L = {\partial M \over \partial H} = {h^{1/\delta -1} \over H_0}
 f_{\chi}(z,z_L)~,
\label{cscale}
\ee
with
\be
f_{\chi}(z,z_L) = {1 \over \delta} f_{G}(z,z_L)
- \frac{z}{\Delta} \frac{\partial f_{G} (z,z_L)}{\partial z}
-\nu_c z_L \frac{\partial f_{G} (z,z_L)}{\partial z_L}~.
\label{fchi}
\ee
A further observable of interest is the thermal susceptibility
$\chi_t$, the mixed second derivative of $f$,
\be
\chi_t = {\partial M \over \partial (1/T)} = -\frac{T^2}{T_0}
h^{(\beta -1)/\Delta}\;\frac{\partial f_{G} (z,z_L)}{\partial z}~.
\label{chit}
\ee
Both $\chi_L$ and $\chi_t$ have their counterparts in QCD.

 We note that our finite-size scaling functions relate in a simple
way to the ones that are commonly used in the literature 
\cite{Mendes,BraunO4}. In terms of these functions our observables
read
\ba
M \;\; & =& \;\; L^{-\beta/\nu}Q_G(z,z_L)~,\label{Mold}\\
\chi_L\;\; & =& \;\; \frac{L^{\gamma/\nu}}{H_0}Q_{\chi}(z,z_L)~,
\label{chiold}\\
\chi_t\;\; & =& \;\; \frac{T^2}{T_0}
L^{-(\beta-1)/\nu}Q_t(z,z_L)~,\label{ctold}
\ea
and the connection to our scaling functions is
\ba
Q_G(z,z_L) \;\; & =& \;\;(z_L/L_0)^{-\beta/\nu}f_G(z,z_L)~,\label{Gold}\\
Q_{\chi}(z,z_L)\;\; & =& \;\;(z_L/L_0) ^{\gamma/\nu}f_{\chi}(z,z_L)~,
\label{Lold}\\
Q_t(z,z_L)\;\; & =& \;\; (z_L/L_0)^{-(\beta-1)/\nu}\left(-
\frac{\partial f_{G} (z,z_L)}{\partial z}\right)~.
\label{told}
\ea
Of particular interest for the determination of the infinite-volume
scaling functions is the leading (in $L$) or asymptotic form of the
$Q$-functions \cite{Mendes}. One obtains for $L\rightarrow\infty$
\be
Q_G\rightarrow f_G(z,0)(hL^{1/\nu_c})^{1/\delta}~,\quad
Q_{\chi}\rightarrow f_{\chi}(z,0)(hL^{1/\nu_c})^{1/\delta-1}~, 
\label{asymp1}
\ee
\be
Q_t\rightarrow -\frac{\partial f_G (z,0)}{\partial z}
(hL^{1/\nu_c})^{(\beta-1)/\Delta}~, 
\label{asymp2}
\ee
that is, at fixed $z$ we have powers of $hL^{1/\nu_c}$
with infinite-volume scaling functions as coefficients. 

\section{\boldmath Simulation of the 3$d$ $O(4)$ model}
\label{section:sio4}

We determine the finite-size scaling functions from simulations of the
standard $O(4)$-invariant nonlinear $\sigma$-model. It is defined by
\be
\beta\,{\cal H}\;=\;-J \,\sum_{<{\vec x},{\vec y}>}\phv_{\vec x}\cdot
\phv_{\vec y} \;-\; {\vec H}\cdot\,\sum_{{\vec x}} \phv_{\vec x} \;,
\ee
where ${\vec x}$ and ${\vec y}$ are nearest-neighbor sites on a
three-dimensional hyper-cubic lattice, $\phv_{\vec x}$ is a
four-component unit vector at site ${\vec x}$. The coupling $J$ and 
the external field $\v H$ are reduced quantities, that is they contain
already  a factor $\beta=1/T$. That allows us to consider the coupling 
directly as the inverse temperature, $J\equiv 1/T$. An additional 
factor on $J$ would just change the scale of $T$. The setup of our 
calculations and the definition of further observables are given in 
detail in Ref.~\cite{Engels2}. There are two differences to the  
simulations in Ref. \cite{Engels:2009tv}, which we used in \cite{Engels2} 
to calculate the infinite-volume scaling functions 
$f(z)\equiv f(z,0)\,$. Instead of calculating at fixed $T$ and 
varying $H$, we fix $z$ and vary $H$ and secondly we use, according
to the average clustersize, an appropriate number of cluster updates such 
as to cover before each measurement the whole volume $V$ of the lattice
5 to 6 times. That becomes relevant in particular with increasing 
$z>0$ and leads there to much better statistics. For example, on a 
lattice of size $L=120$ we made between 400 (at $z=0.1$) and 14.000 
cluster updates (at $z=4.0$) for each of our 100.000 measurements. 
In order to define our variables $t,\bar t,h$ and $z$, we use the same
critical amplitudes, temperature and exponent values for the $3d$ $O(4)$
model as in Refs.\ \cite{Engels:2009tv} and \cite{Fromme}. These are
\be
J_c=0.9359(1),\,\, T_0=1.093(18),\,\, H_0=4.845(66),
\,\, \beta=0.380(2),\,\, \delta=4.824(9)~.
\label{def1}
\ee
From the hyperscaling relations one obtains then
\be
\nu=0.7377(41)~,\quad \gamma=1.4531(104)~,
\quad \Delta=1.8331(103)~, \quad \nu_c=0.4024(2)~.
\label{def2}
\ee
The exponents $\beta$ and $\delta$ were deduced exclusively from Monte
Carlo data for the magnetization at finite external fields.
In Ref.\ \cite{Engels:1999wf} 
the spontaneous magnetization on the coexistence line, that is for
$T<T_c$ and $H=0$, was determined by extrapolating at fixed $T$ from
finite fields to $H=0$, exploiting the Goldstone behavior.
We wanted to convince ourselves that the
value of $\beta$ in Eq.\ (\ref{def1}) is still in agreement with the  
high statistics data for the $3d$ $O(4)$ model that had been produced
on a lattice with $L=120$ at finite $H$ and $J=0.95,0.97,1.0$ and 1.2
for Ref.\ \cite{Engels:2009tv}. We have extrapolated these data also
to $H=0$ in order to improve and to complement the data of 
Ref.\ \cite{Engels:1999wf}. With the complete data set
we have made various fits of the form
\be
     M(T,H=0)=b_M (T_c-T)^{\beta}(1+b_{M1}(T_c-T)^{\theta})~,
\ee
with and without the correction-to-scaling factor, where 
$\theta=0.574$ was taken from Table 3 of the paper
\cite{Guida:1998} by Guida and Zinn-Justin. A free fit with 
corrections resulted in $\beta=0.379(4)$, negligible correction
amplitude $b_{M1}=0.0007(228)$ and $\chi^2/N_f=1.13$. An even better
fit was obtained without corrections, $b_{M1}\equiv 0$, with  
$\chi^2/N_f=0.85$ and $\beta=0.3792(2)$. Further fits, varying 
e.\ g.\ $T_c$ slightly, or omitting the point closest to $T_c$ in
the fits, led us to our error estimate for $\beta$ in Eq.\
(\ref{def1}). Our $\beta$-value differs from the value 
$\beta=0.388(1)$ derived from the exponents $\eta$ and $\nu$ which 
were determined by Hasenbusch \cite{Hasenbusch:2000ph} at 
vanishing external fields on small lattices. A fit to our data with
fixed $\beta=0.388$ without corrections is practically excluded, because
then $\chi^2/N_f=263.5$ and half of the data are outside the fit curve.
Including a correction factor a reasonable fit with $\chi^2/N_f=2.44$
is possible, though at the expense of a noticeable correction
amplitude $b_{M1}=-0.0514(22)$. That however, contradicts another,
well-known result of Ref.\ \cite{Hasenbusch:2000ph}: the leading 
correction-to-scaling amplitude of the $3d$ $O(4)$ nonlinear
$\sigma$-model is essentially zero. That is exactly what we found 
here and in Ref.\ \cite{Engels2}, it was found as well in Ref.\   
\cite{Fromme}, where $\delta$ was determined, and confirmed again
by Ref.\ \cite{Hasenbusch:2011}. We note, that our set of exponents
has been used as well in the successful determination of the scaling
functions of the transverse and even longitudinal (for $z>0$) 
correlation lengths \cite{Engels:2009tv,Fromme}. The exponents, also
our $\nu$-value, are thus consistent with the data. 

The main goal of our simulations was, as already explained, to study 
the finite size dependence of the scaling functions $f_G$ and 
$f_{\chi}$. We have done this by calculating the scaling functions 
at fixed $z$ and varying $z_L=l/h^{\nu_c}$. Here, lattices of size 
$L=24,36,48,72,96$ and external fields $H$ in the range $[0.0005,0.003]$
have been used, supplemented at some $z$-values by results from $L=120$
lattices.
The upper limit of the $H$-range, 0.003, was chosen much smaller than
that of Ref.\ \cite{Mendes}, 0.10, to suppress correction terms
$\sim H^{\omega\nu_c}$. Here, $-\omega=y_4$ is the largest irrelevant
exponent.
The $z$-range that we have considered is $[-1.0,2.0]$, that 
is, we cover the region around the critical point and the high 
temperature region up to and including the peak area of $f_{\chi}(z,0)\,$.
In Ref. \cite{Engels2} we had determined the peak position to 
$z_p=1.374(30)$. It defines the pseudocritical line 
$z=z_p$ in the $(t,h)$-plane where $\chi_L$ is at its maximum for fixed
$h$ and varying $t$.
\begin{figure}[t]
\vspace*{-1.0cm}
\begin{center}
\hspace*{-1.0cm}\includegraphics[width=123mm,angle=-90]{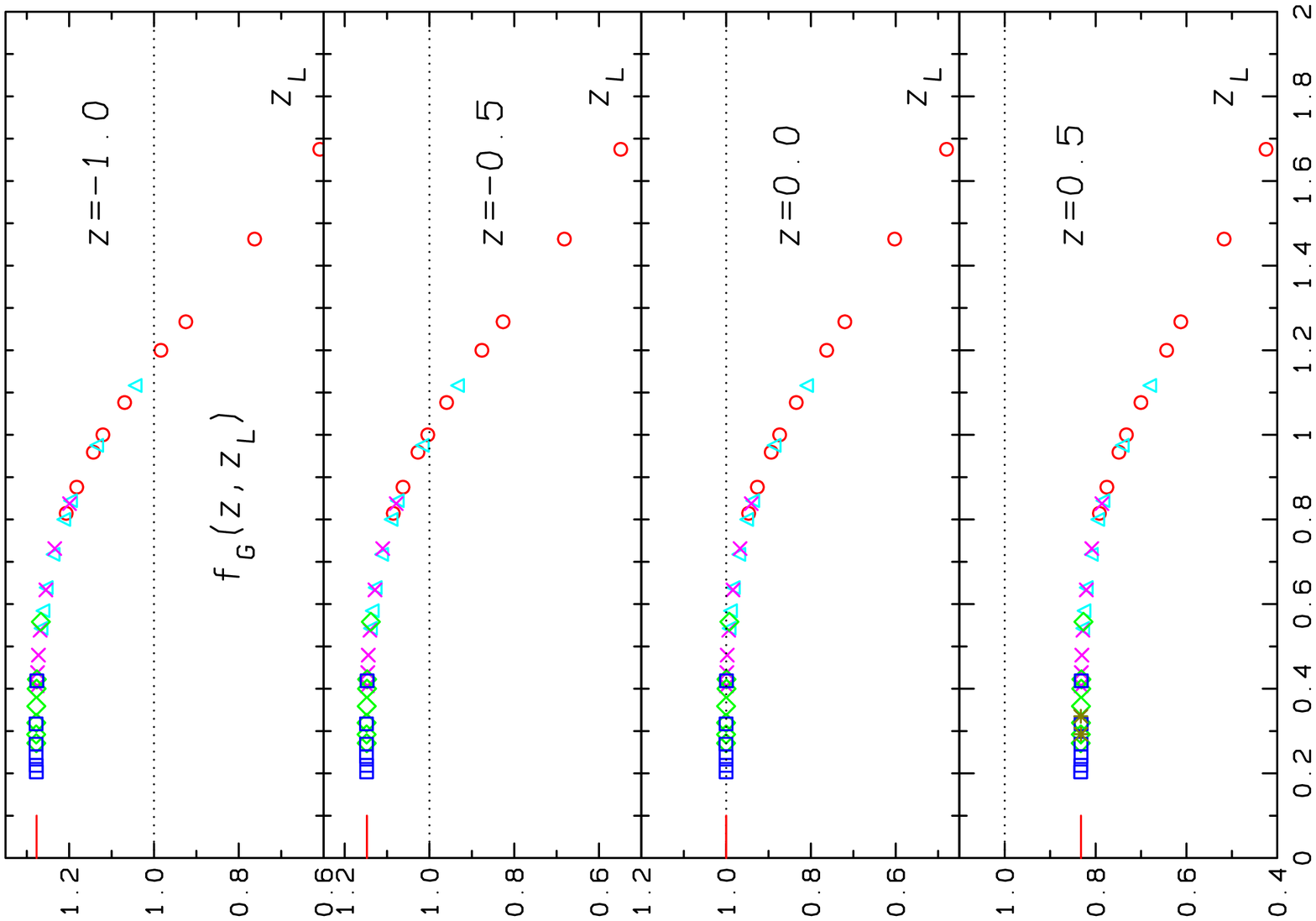}
\hspace*{-1.5cm}\includegraphics[width=123mm,angle=-90]{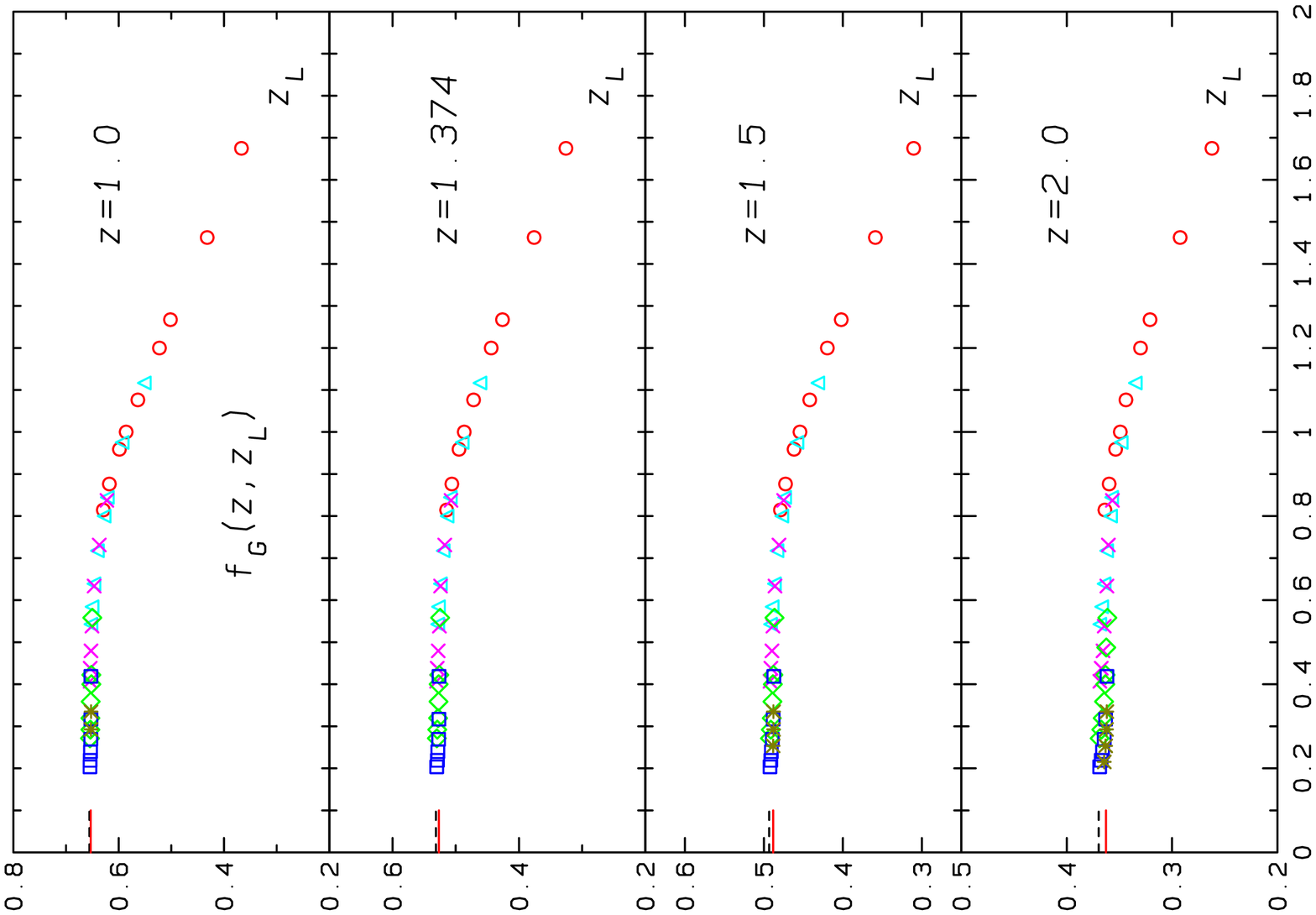}
\hspace*{-0.3cm}
\end{center}
\vspace*{-1.2cm}
\caption{The finite-volume scaling function $f_G(z,z_L)$ for some
values of $z$ versus $z_L$. For each value of $z$ lattices of size
$L=24$ (circles), 36 (triangles), 48 (crosses), 72 (diamonds), 96 
(squares) and various values of $h$ have been used. For 
$z=0.5,1.0,1.5$ and $2.0$ also data for $L=120$ (stars) have been 
added. Short horizontal lines close to $z_L=0$ show the values
of the infinite-volume scaling function $f_G(z)$, the solid lines
denote the new parametrization, the dashed lines the one of Ref.\
\cite{Engels2}.} 
\label{fig:fG_V}
\end{figure}

\begin{figure}[t]
\begin{center}
\includegraphics[width=125mm]{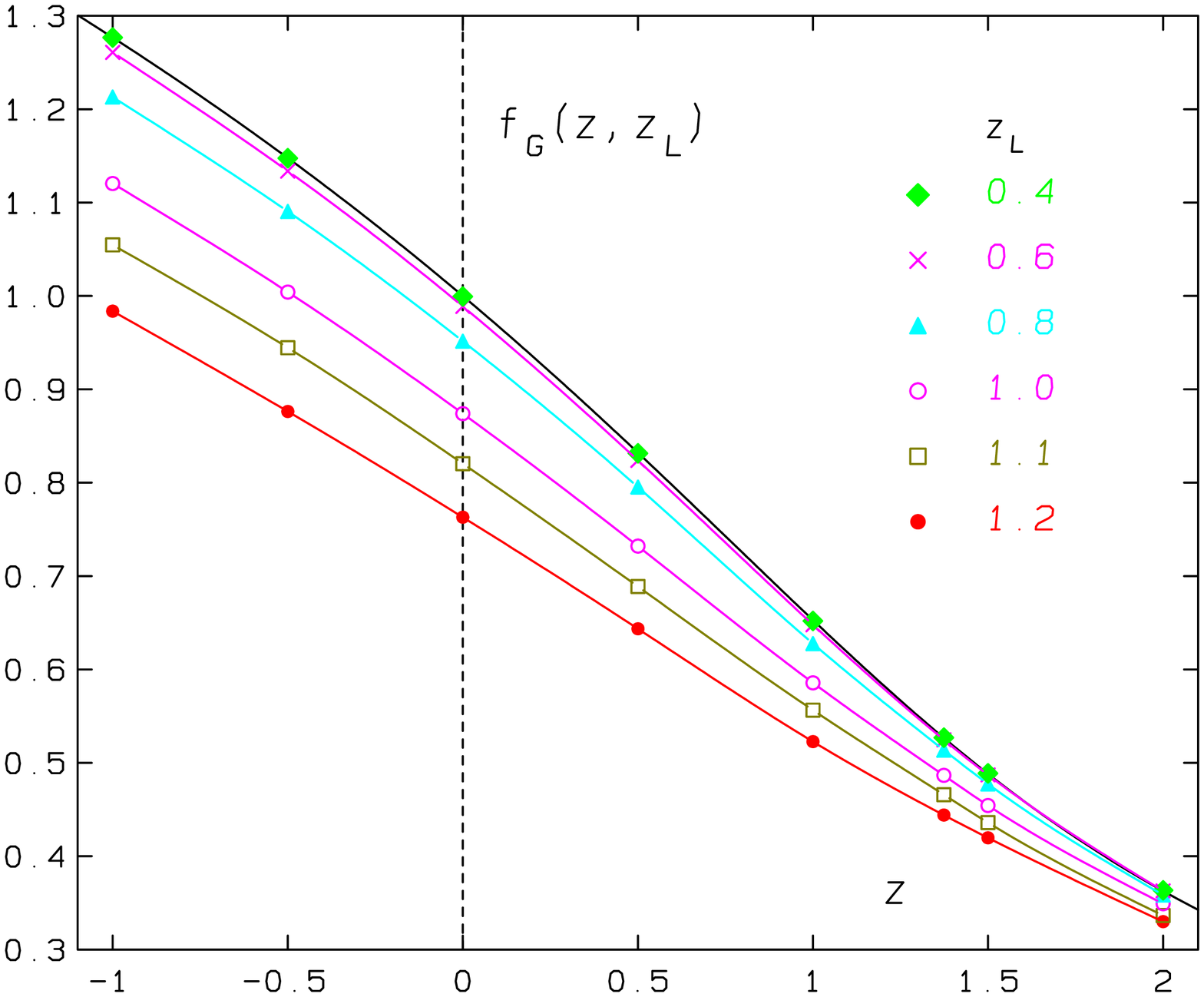}
\end{center}
\vspace*{-0.8cm}
\caption{The finite-volume scaling function $f_G(z,z_L)$ for some
values of $z_L$ versus $z$. Curves shown are spline interpolations
of the data, the points for $z_L=0.4$ are not connected, the highest
line shows $f_G(z,0)$ from the new parametrization.}
\label{fig:fG_V_z}
\end{figure}

All our results for the scaling function $f_G(z,z_L)$ are summarized
in Fig.~\ref{fig:fG_V}. In order to plot the data as a function of 
$z_L=L_0/L h^{\nu_c}$ we have fixed $L_0=1$. As can be seen in the 
$(z=0)$-part of the plot that amounts to the normalization condition 
Eq.\ (\ref{zLnorm}). In  Fig.~\ref{fig:fG_V_z} we have used the same 
data to show the $z$-dependence of $f_G(z,z_L)$ at fixed $z_L$-values.
We see from the two figures that finite size effects in the
scaling function $f_G$ are small for $z_L\le 0.5$. The main effects 
appear in the low temperature region and decrease with increasing $z$.
This is in particular evident from Fig.~\ref{fig:fG_V_z}\,. 

\begin{figure}[ht]
\begin{center}
\includegraphics[width=120mm]{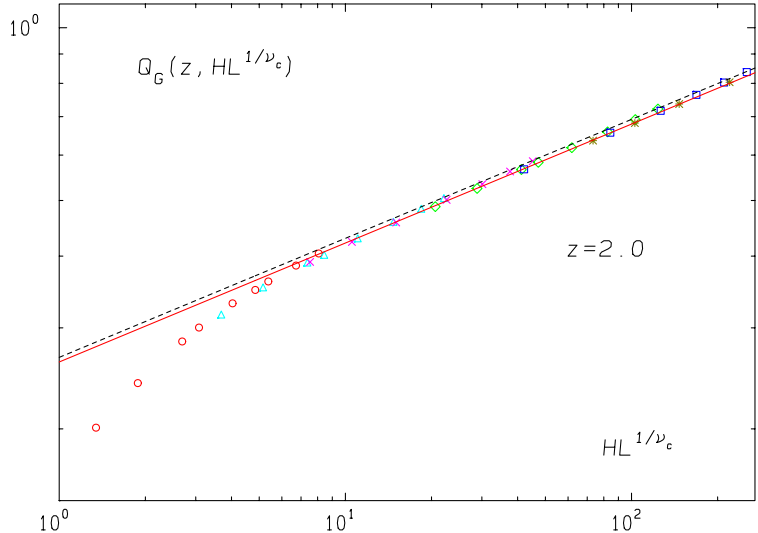}
\end{center}
\vspace*{-0.8cm}
\caption{The finite-volume scaling function $Q_G$ for $z=2$
versus $HL^{1/\nu_c}$. The lines show the asymptotic forms, Eq.\
(\ref{asymp1}), calculated from $f_G(z)$ using the
parametrizations of Ref. \cite{Engels2} (dashed line) and the
one given in Appendix A (solid line).}
\label{fig:QG200}
\end{figure}
\begin{figure}[t]
\vspace*{-1.0cm}
\begin{center}
\hspace*{-1.0cm}\includegraphics[width=123mm,angle=-90]{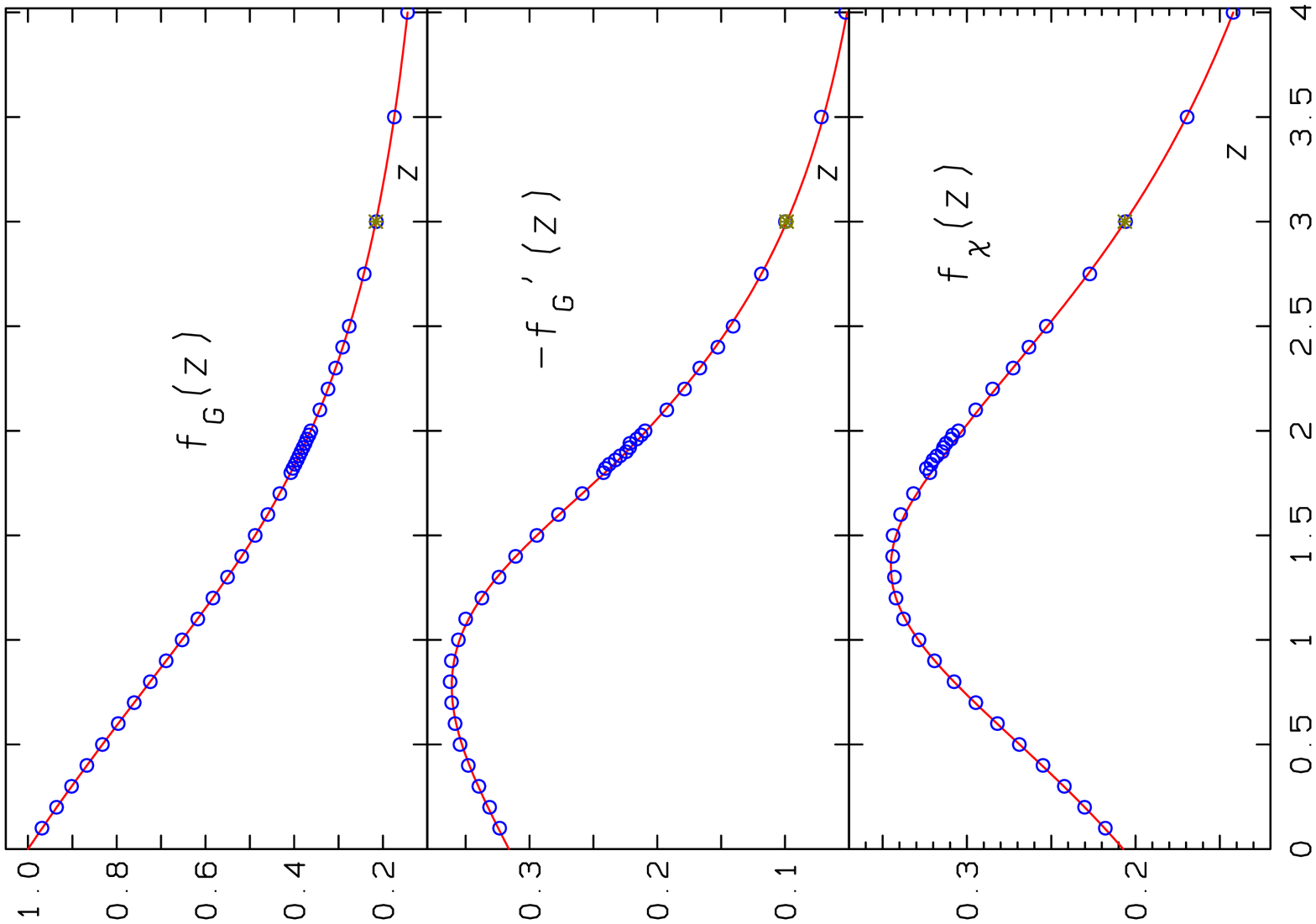}
\hspace*{-1.5cm}\includegraphics[width=123mm,angle=-90]{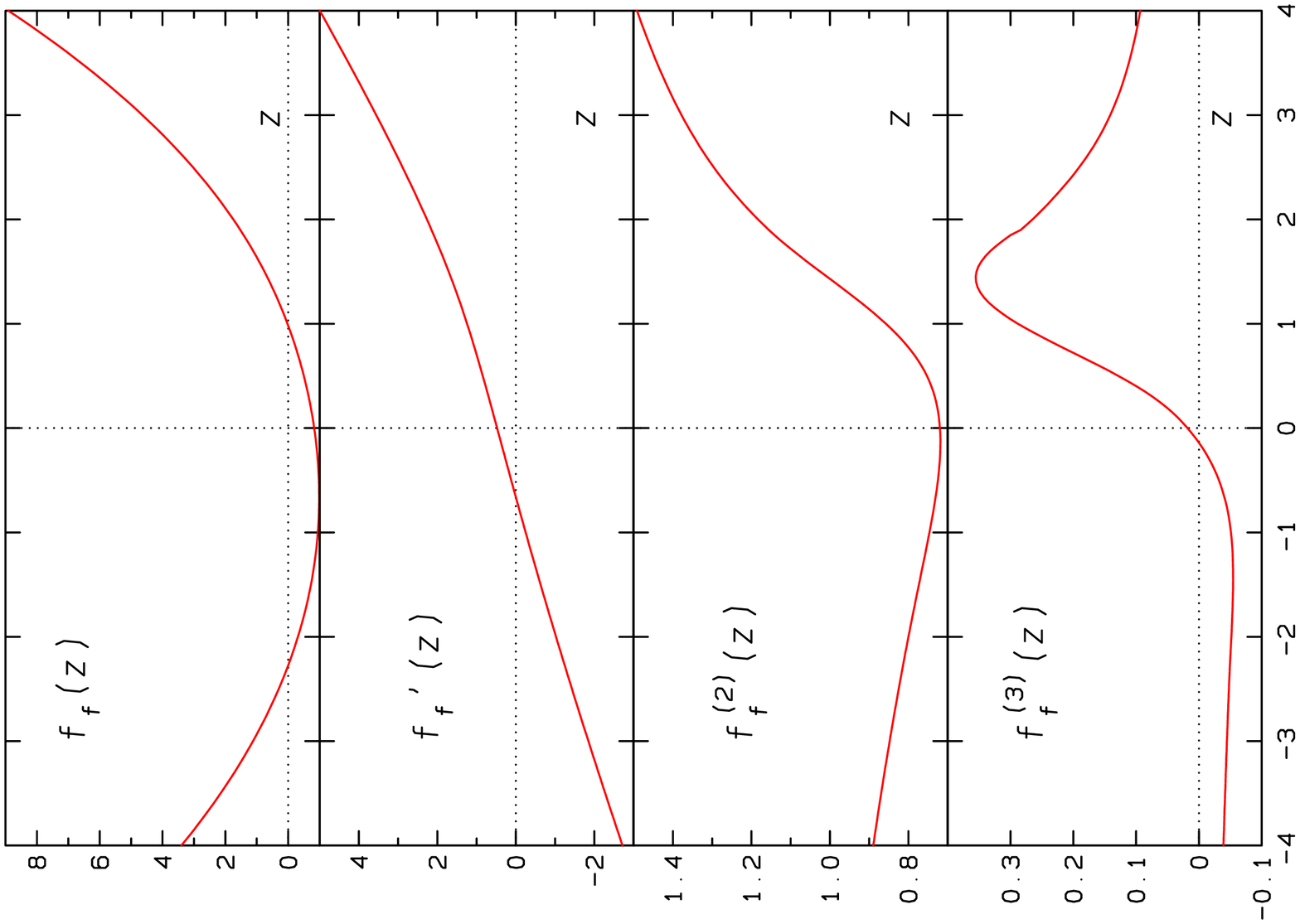}
\hspace*{-0.3cm}
\end{center}
\vspace*{-1.2cm}
\caption{Left part: 
The infinite-volume scaling functions $f_G(z),\,-f_G'(z)$ and 
$f_{\chi}(z)$ for $z>0$, $T>T_c$ from data (circles) for $M,\chi_t$
and $\chi_L$ on an $L=120$ lattice. The lines are from the 
parametrization of the scaling functions in Appendix A. Right part:
The scaling function $f_f(z)$ and its first three derivatives from
the new parametrization.} 
\label{fig:fdata}
\end{figure}
\begin{figure}[t]
\begin{center}
\hspace*{-1.0cm}\includegraphics[width=123mm,angle=-90]{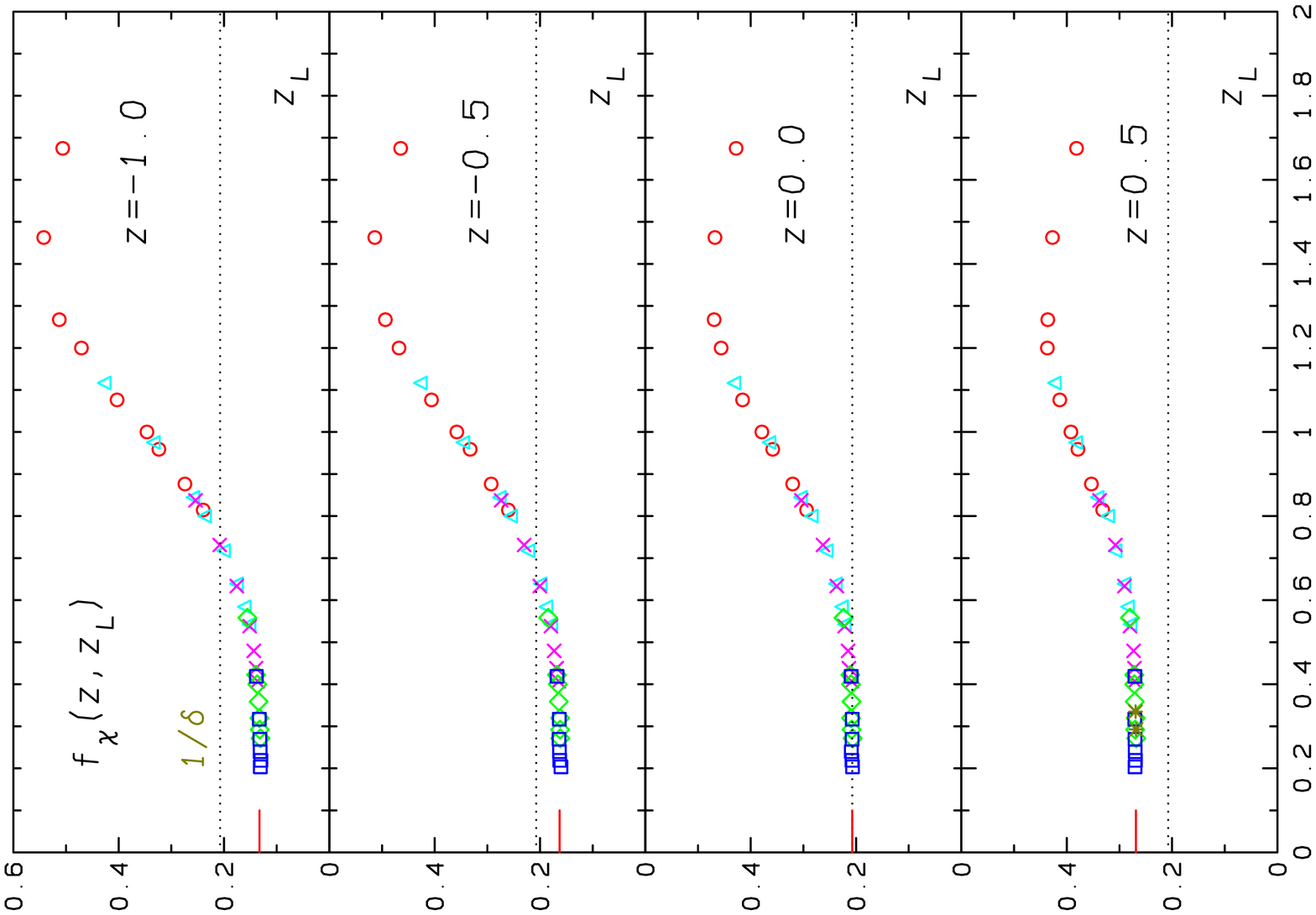}
\hspace*{-1.5cm}\includegraphics[width=123mm,angle=-90]{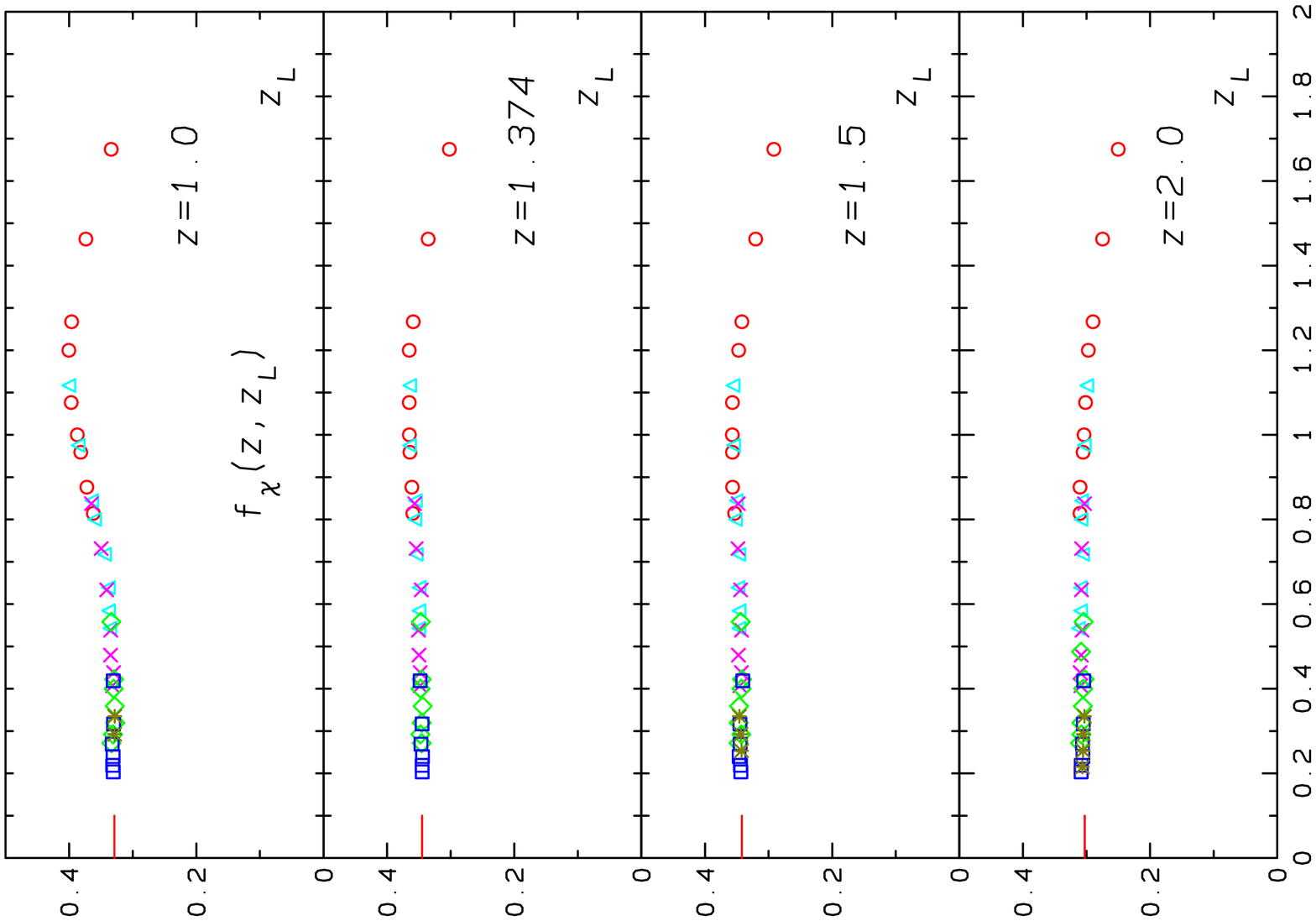}
\hspace*{-0.3cm}
\end{center}
\vspace*{-1.2cm}
\caption{The finite-volume scaling function $f_\chi(z,z_L)$ for some
values of $z$ versus $z_L$. For each value of $z$ lattices of size
$L=24$-$96$ and various values of $h$ have been used. For $z=0.5,1.0,1.5$
and $2.0$ we have also added results from lattices of size $L=120$. The 
notation for the symbols is as in Fig.~\protect\ref{fig:fG_V}\,.  
The infinite volume predictions for $f_{\chi}$ are shown as short 
horizontal lines.}
\label{fig:fchi_V}
\end{figure}
\begin{figure}[t]
\begin{center}
\includegraphics[width=125mm]{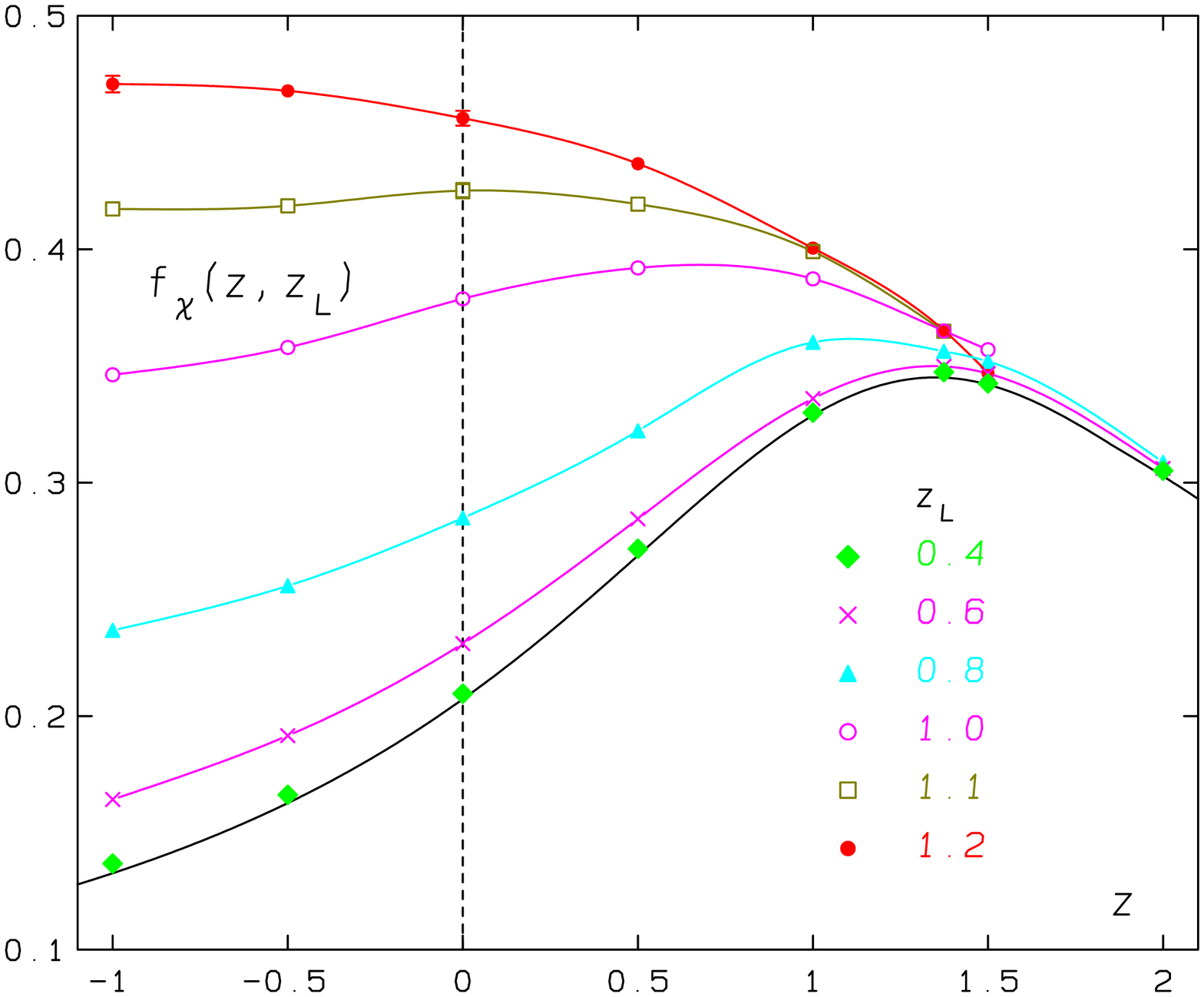}
\end{center}
\vspace*{-0.8cm}
\caption{The finite-volume scaling function $f_\chi(z,z_L)$ for some
values of $z_L$ versus $z$. Curves shown are spline interpolations
of the data, the points for $z_L=0.4$ are not connected, the lowest
line shows $f_{\chi}(z,0)$ from the new parametrization.}
\label{fig:fchi_V_z}
\end{figure}

Another important observation is that we find perfect finite size 
scaling in $z_L$ for $z\le 1.0$. 
That means, at fixed $z$, $f_G$ is de facto only a function of $z_L$
and there is no further significant dependence $\sim H^{\omega\nu_c}$.
At larger values of $z$, however,
the results from different size lattices start to deviate from each 
other. In addition, the data at fixed $z$ do no longer reach a plateau
for $z_L\rightarrow 0$, even for the same lattice size. Obviously,
one needs here data on larger lattices and/or smaller $H$-values. The
effect is known from Ref.\ \cite{Mendes}, where the scaling functions
$Q_G$ had been calculated for $z=0$ and $z=z_p$. In Fig.\ 4 of Ref.\ 
\cite{Mendes}, $Q_G(z_p)$ is shown as a function of $HL^{1/\nu_c}$, 
instead of $z_L$. One observes, that with increasing $H$ the 
corresponding data at fixed $L$ overshoot the asymptotic result
and this happens the earlier the smaller $L$ is. Only for small $H$
and large $L$ there is a small window, where we have coincidence with
the asymptotic behavior expected from Eq.\ (\ref{asymp1}). In fact,
also our parametrization from Ref.\ \cite{Engels2}, which was based 
on results from lattices with $L=120$, is apparently affected by 
the use of data with too large field values in the peak region. In 
order to check this we have made a logarithmic plot of the scaling 
function $Q_G=ML^{\beta/\nu}$ at $z=2$ versus $HL^{1/\nu_c}$ with our
new high statistics data. As is clearly seen in Fig.\ \ref{fig:QG200},
the asymptotic form of $Q_G$ predicted by the parametrization in Ref.\ 
\cite{Engels2} is slightly too high at $z=2$. In order to remedy the
deficiencies we had had in parametrizing the infinite-volume scaling
functions for $z>0$ in Ref.\ \cite{Engels2}, namely too low statistics,
too high field values and no systematic covering of the necessary 
$z$-range, we decided to produce new high statistics data on an $L=120$
lattice at fixed $z$-values with two or three low field values. These 
new data are shown in Fig.\ \ref{fig:fdata} for $f_G(z), -f_G'(z)$ and
$f_{\chi}(z)$ together with a new parametrization, whose details we list 
in Appendix A. The new data cover now in particular the peaks with 
sufficient accuracy. In Figs.\ \ref{fig:fG_V} and \ref{fig:fG_V_z}
we have already displayed the infinite volume values for $f_G$ close
to $z_L=0$. We see that a difference to the old parametrization is here
only visible for $z\gsim 1$. In the right part of Fig.\ \ref{fig:fdata}
we show the scaling function of the free energy density and its first
three derivatives.
Our results for the finite-volume scaling function of the susceptibility
are shown in Figs.~\ref{fig:fchi_V} and \ref{fig:fchi_V_z}. We see
again the strongest finite size effects for the smaller values of $z$.
In particular, the peak of the susceptibility is washed out completely
on small lattices or large $z_L$, so that there a pseudocritical
temperature cannot be determined safely. We note, that the peak position
of $f_{\chi}(z,0)$ has shifted slightly to $z_p=1.35(3)$ due to the new 
parametrization. The value is, however, inside the error bars of the old
value $z_p=1.374(30)$.  
\begin{table}[b]
\begin{center}
\vspace{0.3cm}
\begin{tabular}{|c|c|c|c|c|c|c|}
\hline
$a_{nm}$&~&\multicolumn{5}{|c|}{n}\\
\hline
~&~&0&1&2&3&4\\
\hline
~&3&~0.0421332&-0.0782771&~0.0546495&-0.0251385&~0.0017542\\
~&4&~0.0576183&~0.3302893&-0.2642637&~0.0617961&~0.0049618\\
m&5&-0.6352819&-0.3461722&~0.4678005&-0.0453606&-0.0309722\\
~&6&~0.5355251&~0.1770113&-0.3118316&~0.0061252&~0.0295072\\
~&7&-0.1247180&-0.0369583&~0.0696270&~0.0021488&-0.0078913\\
\hline
\end{tabular}
\end{center}
\caption{Expansion coefficients 
}
\label{tab:anm}
\end{table}
In Figs.~\ref{fig:fG_V_z} and \ref{fig:fchi_V_z} we have shown spline
interpolations of our data. When using the scaling functions in the
analysis of data for other models, it may be more convenient to use 
polynomial interpolation formulas. We have fitted simultaneously all our 
data for $f_G(z,z_L)$ and $f_{\chi}(z,z_L)$ to a polynomial ansatz for
$f_G(z,z_L)$,
\be
f_G(z,z_L) = f_G(z,0) +\sum_{n=0}^4 \sum_{m=3}^{7} a_{nm} z^nz_L^m \; .
\label{ansatz}
\ee

We use here of course the new parametrization of the infinite-volume
scaling function $f_G(z,0)$ as discussed in Appendix A. 
The ansatz for $f_{\chi}(z,z_L)$ is obtained from $f_G(z,z_L)$ using 
Eq.\ (\ref{fchi}). We obtained the expansion coefficients $a_{nm}$ from
a simultaneous fit to data for $f_G(z,z_L)$ and $f_\chi(z,z_L)$.
As can be seen in Figs.~\ref{fig:fG_V} and \ref{fig:fchi_V} 
in the interval $z\in[-1.0,2.0]$ significant finite size effects 
in $f_G$ and $f_\chi$ only set in for $z_L\ge 0.5$. For this reason 
we use in Eq.\ (\ref{ansatz}) a  polynomial ansatz that starts with 
finite volume corrections at ${\cal O}(z_L^3)$. That is, the leading 
corrections are proportional to $1/L^3$. We also performed fits with 
smaller powers of $z_L$ which, however, did not improve over the 
current ansatz. Results for the expansion coefficients $a_{nm}$ are
given in Table~\ref{tab:anm}. This parametrization is appropriate for 
$f_G$ and $f_\chi$ in the intervals $z\in [-1.0,2.0]$ and 
$z_L \in [0.0,1.2]$. 


\section{\boldmath Summary and conclusions}
\label{section:suco}

In our paper we have investigated the finite-size scaling
functions of the universality class of the $3d$ $O(4)$ spin model. Our
aim was to provide a suitable form for tests on this universality 
class of QCD data, which were obtained from simulations on small 
lattices. In contrast to the commonly studied finite-size scaling
functions $Q_G$ \cite{Mendes}, $Q_{\chi}$ and so forth, we have 
directly extended the infinite-volume scaling functions to finite-size 
scaling functions $f(z,z_L)$, which describe the size dependence as well.
The $z$-region, where we have examined these functions, encompasses the
vicinity of the critical point and the high temperature region up to
and including the peak area of $f_{\chi}(z,0)$. Thereby, we cover the
domain where the pseudocritical line obtained from the susceptibility
is of interest. In order to actually use the finite-size scaling 
functions for a test of a model, one has, of course, to determine the
four model specific parameters $T_c, T_0, H_0$ and $L_0$. 

We have seen from Figs.\ \ref{fig:fG_V} and \ref{fig:fG_V_z} for 
$f_G(z,z_L)$ and Figs.\ \ref{fig:fchi_V} and \ref{fig:fchi_V_z} for
$f_{\chi}(z,z_L)$ that on one hand finite size effects are small for
$z_L< 0.5$ and on the other hand that the main finite size effects
appear in the small $z$-region and that they decrease with increasing
$z$. In the course of our analysis we found out that the parametrization
of the infinite-volume scaling function $f_G(z)$ as given in Ref.\ 
\cite{Engels2} had to be improved in the region $z\ge 1.0$\,. We have
therefore generated new high statistics data on an $L=120$ lattice at
fixed $z$ in the range $[0.1,4.0]$ at several small field values each.
These data enabled us to update the parametrization \cite{Engels2}. Its
results are presented in Appendix A. The expansion coefficients given 
in Table \ref{tab:anm} are based on the new parametrization of the 
infinite-volume scaling function.

As a by-product of our calculations we obtained new values for the 
universal product $R_{\chi}=d_0^+=1.0919(14)$ and the universal ratio 
$A^+/A^-=1.734(75)$. Also the peak positions of $f_{\chi}$ and $-f_G'$
changed slightly to $z_{p,\chi}=1.35(3)$ and $z_{p,t}=0.78(4)$,
respectively.
 
\vskip 0.2truecm
\noindent{\Large{\bf Acknowledgment}}

\n This work has been supported in part by contract DE-AC02-98CH10886
with the U.S. Department of Energy.

\appendix
\section{Update on the infinite volume form of \\ the scaling function 
\boldmath$f_G$}
As already shown in Section \ref{section:sio4}, Fig.\  \ref{fig:fdata}, 
we have new data for $z>0$. We have used these data and those of Ref.\
\cite{Engels2} for $z\le 0$ to update the parametrization in \cite{Engels2}
of the infinite-volume scaling functions. We started with the Taylor 
series around $z=0$
\be
f_G(z)\;=\; \sum_{n=0}^\infty b_nz^n~.
\label{sum0}
\ee
Like in \cite{Engels2}, we have fitted the data from $-f_G\p(z)$ for small
$z$. Here, this was done once in the $z$-range
$[-2.5,1.0]$ up to $n=7$, and once in the range $[-0.5,2.3]$ up to $n=8$
but with the coefficients $b_0,\dots,b_4$ from the first fit as input.
The results of the first fit are to be used for $z\le 0$, the others for
$z>0$. The coincidence of the lowest coefficients guarantees smooth
fits at $z=0$ for all the derivatives we need.  
The final result for the coefficients is
\ba
& b_0\equiv 1~,~~b_1=-0.316123\pm 0.000628~,~~b_2=-0.0418371\pm 0.000661~,
\label{b03} \\
& b_3=~0.00129543\pm 0.000859~,~~b_4=~0.00582947\pm 0.000345~.
\label{b35}
\ea
The remaining coefficients are different for $z<0$ and $z>0$. We find
for $z>0$
\ba
 & b_5^+=~0.00201377 \pm 0.000887~,~~b_6^+=~0.0048998\pm 0.001122~,
\label{b56}\\
 & b_7^+=-0.00386218 \pm 0.000484~,~~b_8^+=~0.00068511\pm 0.000071~,
\label{b78}
\ea
and for $z<0$
\ba
 & b_5^-=~0.00257848 \pm 0.000450~,~~b_6^-=~0.00053247\pm 0.000211~,
\label{b56-}\\
 & b_7^-=~0.000044801\pm 0.000030~.
\label{b7-}
\ea

We next consider the asymptotic expansions. In the high temperature region,
that is for $z\rightarrow \infty$, or for $t>0$ and $h\rightarrow 0$,
we use the ansatz
\be
f_G(z)\;=\; z^{-\gamma} \cdot \sum_{n=0}^\infty d_n^+ z^{-2n\Delta}~.
\label{fGas+}
\ee
In the low temperature region, for $t<0$ and $h\rightarrow 0$, 
we make the following ansatz for $f_G(z)$ 
\be
f_G(z)\;=\; (-z)^{\beta} \cdot \sum_{n=0}^\infty d_n^- (-z)^{-n\Delta/2}~.
\label{fGas-}
\ee
We have fitted $f_G$ in the positive $z$-range $[1.5,4]$ with the 
first four terms of Eq.\ (\ref{fGas+}) and found
\ba
& d_0^+=1.09185\pm 0.00137~,~~d_1^+=-1.48536\pm 0.03995~,
\label{d01+}
\\
& d_2^+=3.24559\pm 0.3619~,~~d_3^+=-3.65645\pm 0.9676~.
\label{d23+}
\ea
For negative $z$ we retain the result of the three-term, asymptotic 
fit from \cite{Engels2}
\be
d_0^-\equiv 1~,~~d_1^-=0.273651\pm 0.002933~,~~
d_2^-=0.0036058\pm 0.004875~.
\label{d02-}
\ee
The approximations for the small $z$ and asymptotic expansions, described
above, overlap for both $z>0$ and $z<0$ in large $z$-ranges. For $z>0$ we
change from the small $z$ to the asymptotic fits at $z=1.85$, for $z<0$
at $z=-2.0$. Now everything is fixed and we can calculate the coefficients
of the leading asymptotic terms of $f_f(z)$ from Eqs.\ (59) and (62) of 
Ref.\ \cite{Engels2}. We find
\be
c_0^+ = 0.417756382 \pm 0.00656~,~~ c_0^- = 0.240933670 \pm 0.00913~.
\label{c+-}
\ee
From the last equation one obtains a new estimate of the universal ratio,
\be
\frac{A^+}{A^-}= 1.734 \pm 0.075~.
\label{ratio}
\ee
The new result for the ratio is slightly lower than the results 
$1.842(43)$ from \cite{Engels2}, the estimates found in Refs.\
\cite{ParisenToldin:2003hq}, $1.91(10)$, and \cite{Cucchieri:2004xg},
$1.8(2)$, but still in agreement within the error bars, 
though our error estimate does not include the systematic error
due to a possible variation of the critical exponents used. 

\end{document}